# Misfit phase (BiSe)$_{1.10}$NbSe$_2$ as the origin of superconductivity in niobium-doped bismuth selenide

Machteld E. Kamminga,[a] Maria Batuk,[b] Joke Hadermann[b] and Simon J. Clarke[a*]

[a]Department of Chemistry, University of Oxford, Inorganic Chemistry Laboratory, South Parks Road, Oxford OX1 3QR, United Kingdom.

[b]Electron Microscopy for Materials Science (EMAT), University of Antwerp, Groenenborgerlaan 171, B-2020 Antwerp, Belgium.

[*] E-mail: simon.clarke@chem.ox.ac.uk

**Abstract**

Topological superconductivity is of great contemporary interest and has been proposed in doped $Bi_2Se_3$ in which electron-donating atoms such as Cu, Sr or Nb have been intercalated into the $Bi_2Se_3$ structure. For $Nb_xBi_2Se_3$, with $T_c$ ~3 K, it is assumed in the literature that Nb is inserted in the van der Waals gap. However, in this work an alternative origin for the superconductivity in Nb-doped $Bi_2Se_3$ is established. In contrast to previous reports, it is deduced that Nb intercalation in $Bi_2Se_3$ does not take place. Instead, the superconducting behaviour in samples of nominal composition $Nb_xBi_2Se_3$ results from the $(BiSe)_{1.10}NbSe_2$ misfit phase that is present in the sample as an impurity phase for small $x$ ($0.01 \leq x \leq 0.10$) and as a main phase for large $x$ ($x = 0.50$). The structure of this misfit phase is studied in detail using a combination of X-ray diffraction and transmission electron microscopy techniques.

**Introduction**

Recently there has been significant interest in layered topological insulators such as the chalcogenides $Bi_2Se_3$ and $Bi_2Te_3$.[1–6] Topological insulators are phases of matter in which the bulk material is semiconducting, but the surface contains electrons that are chiral, massless, and conduct electricity as though they were metals. Of particular interest is topological superconductivity, which features the existence of gapless surface states at the surface of a fully gapped semiconductor. Because of their unique electronic structure, topological superconductors are proposed to have great potential in fault-tolerant topological quantum computing.[7] Layered materials often allow for intercalation and/or deintercalation as a way of tuning the chemistry and physics. It has been shown, for example, that intercalation of copper into $Bi_2Se_3$ is possible and gives rise to superconductivity up to 3.8 K.[8,9] Moreover, superconductivity (with $T_c$ ~5.5 K) was observed when $Bi_2Te_3$ was reacted with Pd to form $Pd_xBi_2Te_3$ with $x$ = 0.15, 0.3, 0.5 and 1.[9] Other reported superconducting intercalates are described by the chemical formulae $Sr_xBi_2Se_3$ ($T_c$ ~2.5 K,[7] $T_c$ ~2.9 K[10]) and $Nb_xBi_2Se_3$ ($T_c$ ~3 K[11–13]), where $x$ generally ranges between 0.05 and 0.25. Note that these $M_xBi_2Se_3$ (M = Cu, Sr, Nb) phases are synthesised at elevated temperatures from a stoichiometric ratio of the elements M, Bi and Se. This is in contrast to the work by Koski *et al.*, where zerovalent metals are intercalated, at around ambient temperature, into $Bi_2Se_3$ post synthesis, and complex superstructures as a result of the ordering of the intercalated atoms are obtained.[14]

The chemical nature of these superconducting phases and the location of the intercalated metals in the chalcogenide structure is not trivial. For $Cu_xBi_2Se_3$, synthesised at high temperatures from the elements, Cu is reckoned to be inserted in the van der Waals gap between the $Bi_2Se_3$ quintuple layers,[8,15] as confirmed by the significant increase in the *c*-axis lattice parameter from $c$ = 28.666(1) Å for $Bi_2Se_3$ to $c$ = 28.736(1) Å for $Cu_{0.12}Bi_2Se_3$.[8] A comparable expansion of the unit cell volume is observed by Koski *et al.*[14] using intercalation techniques. On the other hand, for phases described with the formula $Sr_xBi_2Se_3$, there is no clear consensus on the structure. Competing studies state or assume variously that Sr is inserted in the van der Waals gap[10] or within the quintuple layers themselves.[16] For $Nb_xBi_2Se_3$, it was concluded from apparent small shifts in peak positions in powder diffraction measurements that the *c*-axis length expanded slightly leading to the assumption that Nb is inserted in the van der Waals gap of $Bi_2Se_3$ and the compound has a similar structure to $Cu_xBi_2Se_3$.[12,17] The minimal change in lattice parameters implies a highly distorted coordination environment for Nb.[17] Several papers on $Nb_xBi_2Se_3$ describe a range of detailed examinations of the superconducting properties, but in many of these works presented on $Nb_xBi_2Se_3$ samples, the extent of the bulk characterisation of the samples is limited. Kobayashi *et al.*[11] synthesised $Nb_xBi_2Se_3$ with varying $x$ and did examine the samples using bulk powder diffraction measurements. These showed, in addition to a phase with the $Bi_2Se_3$ structure, an impurity phase to which they assigned the formula $BiNbSe_3$[18] and a second impurity, BiSe. These impurities increased in fraction with increasing $x$ and indeed they showed that the phase fraction of the proposed $Nb_xBi_2Se_3$ phase with the $Bi_2Se_3$ structure reaches zero (within the uncertainty) at $x = 0.50$. Work by Wang *et al.*[19] and recent work by Cho *et al.*[20] also show that $Nb_xBi_2Se_3$ samples are multiphase in the bulk. Nevertheless, all these reports assign the superconductivity evident in these samples to a Nb-doped $Bi_2Se_3$-type phase.

Here we show evidence for an alternative explanation for the observed superconductivity in Nb-doped $Bi_2Se_3$ samples of overall composition $Nb_xBi_2Se_3$. We show that conventional high temperature synthesis of samples with the composition $Nb_xBi_2Se_3$ ($0 \leq x \leq 0.50$) results in a mixture of three phases: $Bi_2Se_3$, BiSe and the layered misfit compound $(BiSe)_{1.10}NbSe_2$. This so-called misfit compound is a layered composite (*i.e.* an intergrowth) compound made up of two (or more) interpenetrating sublattices of different chemical composition.[21] For $(BiSe)_{1.10}NbSe_2$, the two sublattices of composition BiSe and $NbSe_2$ occur with a ratio of 1.10 : 1 and have different intralayer lattice constants, which do not match, hence the term *misfit*, and the corresponding misfit ratio is evident in the formula. In this paper and particularly in the supporting information we describe the structure in detail and show how $(BiSe)_{1.10}NbSe_2$ compares with related misfit layered chalcogenides. This misfit phase corresponds to the peaks previously assigned to the $BiNbSe_3$ impurity[18] (which is of fairly similar composition) by Kobayashi *et al.*[11] and by Cho *et al.*[20] We observe that the amount of the phase present with the $Bi_2Se_3$ structure (assumed to be the superconducting phase $Nb_xBi_2Se_3$ in the literature,[11,12,17,19,20,22–26] including very recent literature[19,20,24,25]) in the samples decreases with increasing Nb content in the synthesis mixture and completely vanishes for $x = 0.50$ while the superconducting volume fraction of the sample is directly proportional to the Nb content of the sample.

In addition we do not observe, with synchrotron resolution, a measurable $c$-axis expansion of the $Bi_2Se_3$-structure phase upon Nb doping, suggesting that it is pure and undoped $Bi_2Se_3$. Moreover, our energy-dispersive X-ray spectroscopy (EDX) measurements reveal no crystallites with elemental composition $Nb_xBi_2Se_3$. Therefore, in contrast to previous reports, we deduce that Nb-doping in $Bi_2Se_3$ does not take place and the superconducting behaviour in samples of composition $Nb_xBi_2Se_3$ is actually caused by the misfit phase $(BiSe)_{1.10}NbSe_2$ that is present in the sample as an impurity phase for small $x$ and as a main phase for large $x$.

**Results and discussion**

**Composition dependence of structures and superconducting properties**

Figure 1**a** shows the synchrotron powder diffraction pattern of a sample of composition $Nb_{0.20}Bi_2Se_3$, which is a typical diffraction pattern of these $Nb_xBi_2Se_3$ samples. It is clear that this $Nb_{0.20}Bi_2Se_3$ sample consists of three phases as indicated by the fit and coloured markers: a phase with the $Bi_2Se_3$ structure, the misfit phase $(BiSe)_{1.10}NbSe_2$, and BiSe. The fit for the $Bi_2Se_3$ structure phase resulted in lattice parameters of $a$ = 4.1390(8) Å and $c$ = 28.6298(10) Å which are equal within one estimated standard deviation in the refinement to those of pure $Bi_2Se_3$ ($a$ = 4.1393(11) Å and $c$ = 28.6304(9) Å) which we measured with synchrotron radiation under the same conditions, from which we conclude that there is no evidence for Nb intercalation. Figures 1**c-e** show representative scanning electron microscopy (SEM) energy dispersive X-ray spectroscopy (EDX) maps of the sample. Figure 1**b** shows crystallites with different elemental composition corresponding to either $(BiSe)_{1.10}NbSe_2$ with a Nb:Bi ratio close to 1:1 or Nb-free Bi and Se phases, *i.e.* no crystallites with a composition close to $Nb_xBi_2Se_3$ are observed for this sample with $x$ = 0.2 in the reaction mixture. The chemical composition of the misfit phase was determined by SEM-EDX as Bi/Se/Nb: 22(1)/59(1)/19(1) (Fig. 1) and additionally by scanning transmission electron microscopy (STEM) EDX measurements as Bi/Se/Nb: 21.4(2)/60.5(8)/18.1(7) (Supplementary Fig. S10), which are in agreement within the experimental uncertainty. Moreover Nb is only found in combination with Bi and Se in a ratio corresponding to $(BiSe)_{1.10}NbSe_2$, no other Nb-containing phase such as $NbSe_2$ is observed.

In Fig. 2**a**, we show the superconducting properties of samples made with compositions $Nb_xBi_2Se_3$. The critical temperature $T_c$ of roughly 3.2 K is in agreement with previous reports of superconductivity in samples of these nominal compositions.[11–13] Moreover, the superconducting volume fraction is directly proportional to $x$. We state the susceptibility per gram of material due to the fact that the $Nb_xBi_2Se_3$ sample is a mixture of three phases and the ratios of these phases differ for different $x$. Figure 2**b** shows the first two characteristic peaks of the diffraction pattern: the 003 peak of the $Bi_2Se_3$-structure phase and the 002 peak of $(BiSe)_{1.10}NbSe_2$. The intensity of the $Bi_2Se_3$-structure 003 peak is reduced upon increasing $x$ and for $x = 0.50$ there is no evidence for a $Bi_2Se_3$-structure phase present in the sample. Moreover, as quantified above, we do not observe any evidence, with synchrotron resolution, for a measurable peak shift and hence a $c$-axis expansion within one estimated standard deviation of this $Bi_2Se_3$-structure phase present in the samples upon Nb doping (see Fig. 2**c**) suggesting that it is undoped $Bi_2Se_3$. This is in contrast to the variations stated in literature reports. However these reports either do not refine the lattice parameters from the whole pattern,[11] or do not present any uncertainties on the refined values,[19] or do not make a comparison using the same measurement conditions with a pure $Bi_2Se_3$ phase.[20] Wang[19] report a slight expansion in $c$ on Nb doping, but do not report the estimated standard deviation and Cho et al.[20] report a $c$ lattice parameter of 28.4633(7) Å, which is much smaller than what we and Wang et al.[19] find for pure $Bi_2Se_3$. Our previous work on $Li_{1-x}Fe_{1+x}As$ samples showed that trends in lattice parameters arising from very small compositional changes could only be monitored by synchrotron radiation.[27] Our observation of no $c$-axis expansion within one estimated standard deviation with synchrotron resolution is strong evidence that no intercalation of Nb into $Bi_2Se_3$ took place. This argument, in combination with the fact that the superconducting volume fraction increases with $x$ while the amount of the $Bi_2Se_3$-structure phase decreases with $x$, proves that the superconductivity does not arise from a $Bi_2Se_3$-structure phase. Thus, the superconductivity does not arise from the assumed intercalate $Nb_xBi_2Se_3$ phase with Nb intercalated between or within $Bi_2Se_3$ slabs or substituted for one of the other elements. Particularly compelling is that the superconducting volume fraction is highest when there is no $Bi_2Se_3$-structure phase present within the detection limits of the powder diffraction experiment. In addition, the intensity of the 002 peak of $(BiSe)_{1.10}NbSe_2$ is directly proportional to the superconducting volume fraction (Fig. 2**d**). For small $x$, (BiSe)$_{1.10}$NbSe$_2$ can be considered a small superconducting impurity and for larger $x$, the (BiSe)$_{1.10}$NbSe$_2$ phase dominates. Note that Fig. 2**a** shows that there appears to be a slight variation in $T_c$ (between 3.0 K and 3.2 K) among samples prepared under similar conditions with various Nb contents. As will be highlighted below, (BiSe)$_{1.10}$NbSe$_2$ is subject to small variations in $T_c$ depending on synthesis technique, and can contain various stacking faults.

As shown in Fig. 1, BiSe is also present in Nb$_x$Bi$_2$Se$_3$ samples in addition to Bi$_2$Se$_3$ and (BiSe)$_{1.10}$NbSe$_2$. As can be seen in Fig. 2**b**, when $x$ reaches 0.5, Nb$_{0.50}$Bi$_2$Se$_3$ contains no phase with the structure of Bi$_2$Se$_3$ and therefore solely consists of the misfit phase (BiSe)$_{1.10}$NbSe$_2$ and BiSe according to the equation:

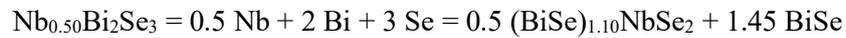

$$\text{Nb}_{0.50}\text{Bi}_2\text{Se}_3 = 0.5\ \text{Nb} + 2\ \text{Bi} + 3\ \text{Se} = 0.5\ (\text{BiSe})_{1.10}\text{NbSe}_2 + 1.45\ \text{BiSe}$$

Our investigation of the relationship between the superconducting volume fraction and the ratio of (BiSe)$_{1.10}$NbSe$_2$ and BiSe in Nb$_{0.50}$Bi$_2$Se$_3$ reveals that the superconducting volume fraction increases with the (BiSe)$_{1.10}$NbSe$_2$/BiSe ratio, confirming that the misfit phase is responsible for the superconducting properties in samples of composition Nb$_x$Bi$_2$Se$_3$ and BiSe is not. A full description of this investigation is given in the Supplementary Section 1 and Supplementary Fig. S5.

Figure 3 shows the magnetometry data for the sample of composition Nb$_{0.50}$Bi$_2$Se$_3$ in comparison with the superconducting properties of its two constituents *i.e.* BiSe and the misfit phase (BiSe)$_{1.10}$NbSe$_2$ synthesised in pure form (see Methods section). The pure BiSe sample is not superconducting while the pure misfit phase is superconducting with a similar $T_c$ as that found for the mixed-phase sample of composition Nb$_{0.50}$Bi$_2$Se$_3$ as shown in Fig. 2. Note that bulk 2H-NbSe$_2$ is a well-studied superconductor with a $T_c \sim 7$ K.[28–31] Monolayer NbSe$_2$ has a $T_c \sim 3$ K,[32] which is similar to (BiSe)$_{1.10}$NbSe$_2$. This is consistent with the misfit phase being effectively constructed of single layers of NbSe$_2$ separated by the BiSe layers. Since the scanning electron microscopy (SEM) data of Supplementary Fig. S4 reveal that this Nb$_{0.50}$Bi$_2$Se$_3$ sample contains no crystallites with the Nb$_x$Bi$_2$Se$_3$ composition and solely consists of BiSe and (BiSe)$_{1.10}$NbSe$_2$, as confirmed by the synchrotron X-ray diffraction (XRD) data from Fig. 2, this data directly proves that the superconductivity in Nb$_{0.50}$Bi$_2$Se$_3$ and therefore in all our Nb$_x$Bi$_2$Se$_3$ samples is caused by the superconducting (BiSe)$_{1.10}$NbSe$_2$ misfit compound.

Several papers report the presence of BiNbSe$_3$ impurities, but do not regard these impurities as relevant for the reported physical properties.[11,19,20] Wang et al.,[19] attempted to make this impurity as a pure phase, but no structure or physical properties were given. Our diffraction and electron microscopy measurements, as discussed below, show that the peaks assigned to the BiNbSe$_3$ phase[18] could actually be assigned to the superconducting (BiSe)$_{1.10}$NbSe$_2$ misfit phase. Thus, to the best of our knowledge, no single phase of alleged Nb$_x$Bi$_2$Se$_3$ nor direct proof of Nb intercalation into Bi$_2$Se$_3$ by means of single-crystal XRD, electron diffraction or electron microscopy imaging techniques is reported to date.

A large portion of the published work on the interesting physical properties reported for Nb$_x$Bi$_2$Se$_3$ are performed on single crystals,[12,13,20,22,33,34] that are provided by a single research group who described their synthesis method in Ref.[17] The most recent of these papers by Cho et al.,[20] performed bulk analysis on a sample of nominal composition Nb$_{0.25}$Bi$_2$Se$_3$ in response to the published referee report. According to the published powder diffraction pattern, that analysis showed that the sample used was a similar multiphase mixture to those which we report here and there is also consistency with the work of Kobayashi et al.[11] Thus, all the published works which show bulk analysis of Nb$_x$Bi$_2$Se$_3$ samples show multiphase behaviour, and the published works on single crystals mentioned above have studied single crystals that were selected from these multiphase batches.

As shown in Supplementary Fig. S3, we found a small variation in T$_c$ of (BiSe)$_{1.10}$NbSe$_2$ depending on the synthesis technique. Herein, we show the magnetometry data of naturally cooled (*i.e.* at the rate of the furnace when switched off) and ice/water quenched pure misfit samples as obtained by the vapour transport method.[35] The naturally cooled sample has a T$_c$ ~2.3 K which is in good agreement with previous reports on the superconducting properties of the misfit phase.[36] The ice/water quenched sample has a significantly higher T$_c$ that is comparable to the ice/water quenched samples of Nb$_x$Bi$_2$Se$_3$ shown in Fig. 2**a**. We do not observe any notable difference in the X-ray patterns between the samples. The (BiSe)$_{1.10}$NbSe$_2$ data in Fig. 3 is obtained by annealing the ice/water quenched sample at 640 °C for a couple of days. We hypothesise that the small changes in properties are caused by either minuscule compositional changes (e.g. Se deficiencies) or potentially different NbSe$_2$ stacking faults (see below) that can be present in the (BiSe)$_{1.10}$NbSe$_2$ compound. This study will be part of future investigations.

**Structural analysis of (BiSe)$_{1.10}$NbSe$_2$**

(BiSe)$_{1.10}$NbSe$_2$ has been synthesised and studied previously,[35,36] but complete structure solution by single-crystal X-ray diffraction was hampered, apparently by disorder in the BiSe sublattice along the stacking direction. To probe the structure of the misfit phase (BiSe)$_{1.10}$NbSe$_2$ in more detail we synthesized the pure phase from the elements using a vapour transport synthesis method.[35] Powder diffraction data (Fig. 4a) are in agreement with those of Zhou et al.[35] Single-crystal X-ray diffraction measurements performed on the I19 beamline at Diamond Light Source (UK) show severely streaked reflections parallel to c (Supplementary Fig. S6), confirming a type of disorder intrinsic to this phase and restricting the amount of structural information that can be obtained from single-crystal X-ray diffraction. Refinement of composite structures from powder diffraction data,[37–39] gives less detailed models than those obtained from single-crystal X-ray diffraction[38] as there is no reliable method to determine a modulation vector from powder data.[37] Therefore, as the starting point for refining the crystal structure from synchrotron powder X-ray data, we performed transmission electron microscopy and combined its results with the previously published model for (BiSe)$_{1.09}$TaSe$_2$,[35,40] which is very similar to (BiSe)$_{1.10}$NbSe$_2$[35]. A detailed description of the modelling of the average crystal structure of (BiSe)$_{1.10}$NbSe$_2$ (summarised in Figures 4b,c) from synchrotron powder data and electron diffraction data is given in the Supplementary Section 2 and Supplementary Figures S8 and S9. In this description the crystal consists of two separate translationally symmetric subsystems. Due to their mutual interaction, the subsystems in the real crystal structure will be modulated.[41]

In order to gain more insight into the local structure of (BiSe)$_{1.10}$NbSe$_2$ and how it differs from that of the Ta analogue, we performed high angle annular dark field (HAADF) scanning transmission electron microscopy (STEM) imaging. Figures 5a,b show representative HAADF-STEM images acquired along the [100] direction. They clearly show a layered structure with alternating BiSe and NbSe$_2$ blocks. The thickness of the blocks and their [100] orientation were constant in the whole area of investigation. The synthesis technique used here leads to a more uniform structure than the layer-by-layer vapour deposition employed by Mitchson et al.[42,43] The structure of $(BiSe)_{1.10}NbSe_2$, refined (Fig. 4a) using the $(BiSe)_{1.09}TaSe_2$[35,40] model as a starting point, is overlaid on the images. The image in Fig. 5a perfectly matches the refined structure from Fig. 4b,c and all $NbSe_2$ layers have the same orientation (as indicated by the green triangles which indicate the orientation of the $NbSe_6$ triangular prisms). However, in some regions of the crystals some of the $NbSe_2$ layers have two different orientations, and the two orientations are related by a reflection in the *ac* plane. The second orientation is represented by the red triangles. The presence of the two orientations can be seen from the shifts of Nb and Se atomic columns in Fig. 5b and as depicted in the figure legend. The two types of $NbSe_2$ layers with opposing orientations of the $NbSe_6$ triangular prisms can perfectly alternate along the stacking direction (Fig. 5b) or occur in a random manner (Fig. 5c), obtaining the character of stacking faults. The corresponding electron diffraction pattern acquired from a large area is shown in Fig. 5d. It can be indexed with the cell parameters $b \approx 6.0$ Å, $c \approx 24.0$ Å, in agreement with the refinement parameters in Supplementary Tables S1 and S2, but shows elongated reflections $0kl$ when $l$ is odd, which is in agreement with the occurrence of these stacking faults and explains the streaked reflections observed for single-crystal X-ray diffraction in Supplementary Fig. S6 and as described in reference 35 which accounts for the difficulty in refining the structure from single-crystal data, and is a difference between $(BiSe)_{1.10}NbSe_2$ and the Ta analogue. Further explanation of how these stacking faults observed by STEM relate to the powder pattern as shown in Fig. 4a is given in the Supplementary Section 3, Supplementary Figures S10-S12 and Supplementary Table S3.

**Concluding remarks**

In conclusion, there are several reports in the literature of derivatives of the important narrow band gap semiconductor $Bi_2Se_3$ which show superconductivity. A notable contemporary example is the case of samples with the proposed formula $Nb_xBi_2Se_3$. In this work we show that there is no evidence for the chemically surprising formation of phases in which Nb is intercalated between the $Bi_2Se_3$ quintuple layer slabs or otherwise inserted or substituted into the $Bi_2Se_3$ structure. We conclude from analysis of high resolution powder diffraction data, transmission electron microscopy and magnetometry measurements on samples of overall composition $Nb_xBi_2Se_3$ ($0 \leq x \leq 0.50$) that pure $Nb_xBi_2Se_3$ phases with the $Bi_2Se_3$ structure are not attainable using high temperature synthetic methods, and that the superconductivity in these compositions arises from the misfit phase $(BiSe)_{1.10}NbSe_2$ which is a composite of $NbSe_2$ and $BiSe$ slabs. This evident inhomogeneity of samples of nominal composition $Nb_xBi_2Se_3$ might also explain why in one report it was noted that not all of the samples were of high enough quality to show a large superconducting volume fraction.[33] Furthermore the upper critical field data are reportedly dominated by 2-fold symmetry, in contradiction to the 3-fold symmetry of the $Bi_2Se_3$ structure which led to the proposal of a nematic state.[22] We suggest that it should be checked whether the orthorhombic two-fold symmetry of the superconducting misfit phase $(BiSe)_{1.10}NbSe_2$ can account for the upper critical field response.

**Methods**

### Synthesis

Polycrystalline samples of composition $Nb_xBi_2Se_3$ ($x$ = 0.01, 0.05, 0.10, 0.20, 0.30, 0.40, 0.50) were synthesized by mixing together stoichiometric ratios of high-purity Bi pieces (Sigma Aldrich; 99.999%), Se powder (Alfa Aeser; 99.999%) and Nb powder (Alfa Aeser; 99.99%), and heating them in an evacuated sealed silica tube to 850 °C at 30 °C h$^{-1}$. After holding the temperature for 72 h, the powders were cooled slowly to 610 °C at 3 °C h$^{-1}$ and finally quenched into ice/water after which the obtained products were ground into a fine powder. This procedure is closely related to a previously reported procedure for the synthesis of $Sr_xBi_2Se_3$,[7] and was chosen as it was very similar to various reports on the synthesis of $Nb_xBi_2Se_3$,[11,25] in order to compare the results. The very slow cooling between 850 °C and 610 °C was performed in order to allow crystals to grow from the melt.[11,17,19,25] Ice/water quenching was performed as this was done in previous reports[11,19,25] with which we sought comparison, and it was found to be crucial for obtaining superconductivity in $Sr_xBi_2Se_3$.[16] To verify the need for ice/water quenching, we synthesised single crystals of $(BiSe)_{1.10}NbSe_2$ by a vapour transport method with and without ice/water quenching as detailed below.

Polycrystalline BiSe was synthesised by mixing together stoichiometric ratios of the Bi pieces and Se powder described above, following a previously reported synthesis method.[44] The elements were heated in an evacuated sealed silica tube to 850 °C at 2 °C min$^{-1}$. After holding the temperature for 1 week, the samples were quenched into ice/water, ground into a fine powder and annealed at 550 °C in order to increase crystallinity for roughly 2 months followed by quenching in ice/water.

Single crystals of the layered misfit compound (BiSe)$_{1.10}$NbSe$_2$ were synthesized through a previously reported vapour transport method.[35] A stoichiometric ground mixture of the elements described above was placed in one end of an evacuated sealed silica tube and this end of the tube was placed in a tube furnace at 700 °C with the other end maintained at 640 °C. Around 10 mg of (NH$_4$)$_2$PbCl$_6$ per 500 mg of total mass of the precursors was used as transport agent. (NH$_4$)$_2$PbCl$_6$ was prepared by a modified synthesis method[45] from a stoichiometric mixture of PbO$_2$ and NH$_4$ in ice cold HCl. After around 10 days, the tube was ice/water quenched and the crystals were extracted from the cold end of the tube. Small crystals with sizes up to 0.5 mm were obtained. Naturally cooled samples without ice/water quenching (*i.e.* cooled at the rate of the tube furnace when switched off) were also prepared and the structures and properties of the obtained crystals were compared.

**X-ray diffraction**

Powder X-ray diffraction (XRD) measurements were performed on both laboratory XRD equipment (Bruker D8 Advance, Cu Kα radiation) and on the I11 beamline at the Diamond Light Source (UK). At I11, diffractograms were measured upon exposure of the monochromatic 15.0 keV (λ = 0.82655(1) or 0.82640(8) Å calibrated using a Si standard) X-ray beam and collected using a Mythen position sensitive detector. Single-crystal XRD measurements were performed on the I19 beamline at Diamond Light Source (UK). Data analysis was performed with TOPAS-Academic V5[46] and Jana2006.[47]

**Electron microscopy**

Scanning electron microscopy (SEM) measurements were performed using a FEI Quantum FEG 650 operating in low vacuum mode with an accelerator voltage of 20 kV and a spot size of 3.5. Electron diffraction patterns were acquired on a Thermo Fisher Tecnai transmission electron microscope (TEM) operated at 200 kV. High angle annular dark field (HAADF) scanning transmission electron microscopy (STEM) images and energy dispersive X-ray spectroscopy (EDX) maps were acquired using a Thermo Fisher Titan 80-300 "cubed" microscope operated at 300 kV. Specimens for the TEM study were prepared in two ways: (1) the material was dispersed in ethanol for 1 h using an ultrasonic bath, with regular breaks to allow the water in the bath to cool. Then, a few drops of the suspension were deposited onto a copper TEM grid covered by a holey carbon layer. (2) The cross-section and plane-view lamellae were prepared from individual crystals using focused ion beam (FIB) milling. During the preparation process Pt and carbon protective layers were deposited on top of the film. The specimens were prepared in air. TEM images obtained for the specimens prepared by the different methods were consistent. For the compositional analysis by STEM-EDX, Bi-M, Nb-L and Se-K lines were used.

### Magnetometry measurements

Magnetic susceptibility measurements were conducted using a Quantum Design MPMS3 superconducting quantum interference device (SQUID) magnetometer. Gelatin capsules were used to contain accurately-weighed powder samples of about 20-30 mg in mass. Measurements were performed on warming in a d.c. field of 10 Oe in the temperature range 1.8-10 K after firstly cooling in zero applied field (ZFC) and then after cooling in the applied field of 10 Oe (FC).

**Acknowledgements**

M.E.K. was supported by the Netherlands Organisation for Scientific Research (NWO, grant code 019.181EN.003). We also acknowledge support from the EPSRC (EP/R042594/1, EP/P018874/1, EP/M020517/1). J.H. acknowledges support from the University of Antwerp through BOF Grant No. 31445. We thank DLS Ltd for beam time (EE18786), Dr Clare Murray for assistance on I11 and Dr Jon Wade from the Department of Earth Sciences, University of Oxford for performing the SEM measurements. We also thank Dr Michal Dušak and Dr Václav Petříček for their advice concerning the use of the Jana2006 software.

**Author contributions**

M.E.K. and S.J.C. synthesised the compounds, performed the diffraction and magnetometry experiments, and performed the structural refinement. M.B. and J.H. performed the TEM and HAADF-STEM measurements and corresponding analysis. M.E.K. wrote the paper involving all authors.

**Competing interests**

The authors declare no competing interests.

**Additional information**

Supplementary information is available for this paper at the end of this document.

Correspondence and requests for materials should be addressed to S.J.C.

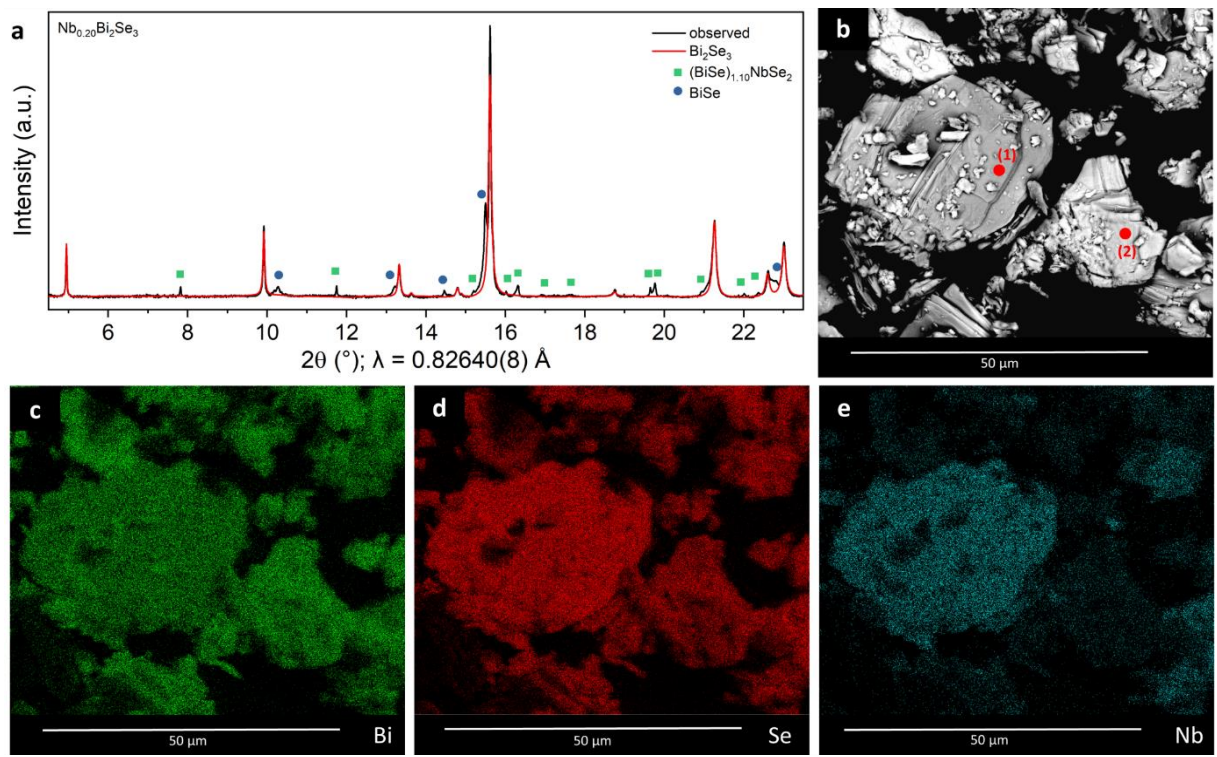

**Fig. 1 Diffraction and electron microscopy data of Nb$_{0.20}$Bi$_2$Se$_3$. a** Synchrotron X-ray diffraction pattern of Nb$_{0.20}$Bi$_2$Se$_3$ fitted with Bi$_2$Se$_3$. Green squares and blue dots indicate impurity phases of the misfit phase (BiSe)$_{1.10}$NbSe$_2$ and BiSe, respectively. **b** Electron Backscatter Diffraction (EBSD) image of the same sample showing crystallites with different average $Z$ value. Point (1) has an elemental ratio of ~1 Nb : ~1.1 Bi : ~3 Se, which closely resembles (BiSe)$_{1.10}$NbSe$_2$,[35] whereas point (2) has an elemental ratio of ~2 Bi : ~3.4 Se with no Nb present. **c-e** EDX maps corresponding to **b** showing the presence of Bi, Se and Nb in the crystallites.

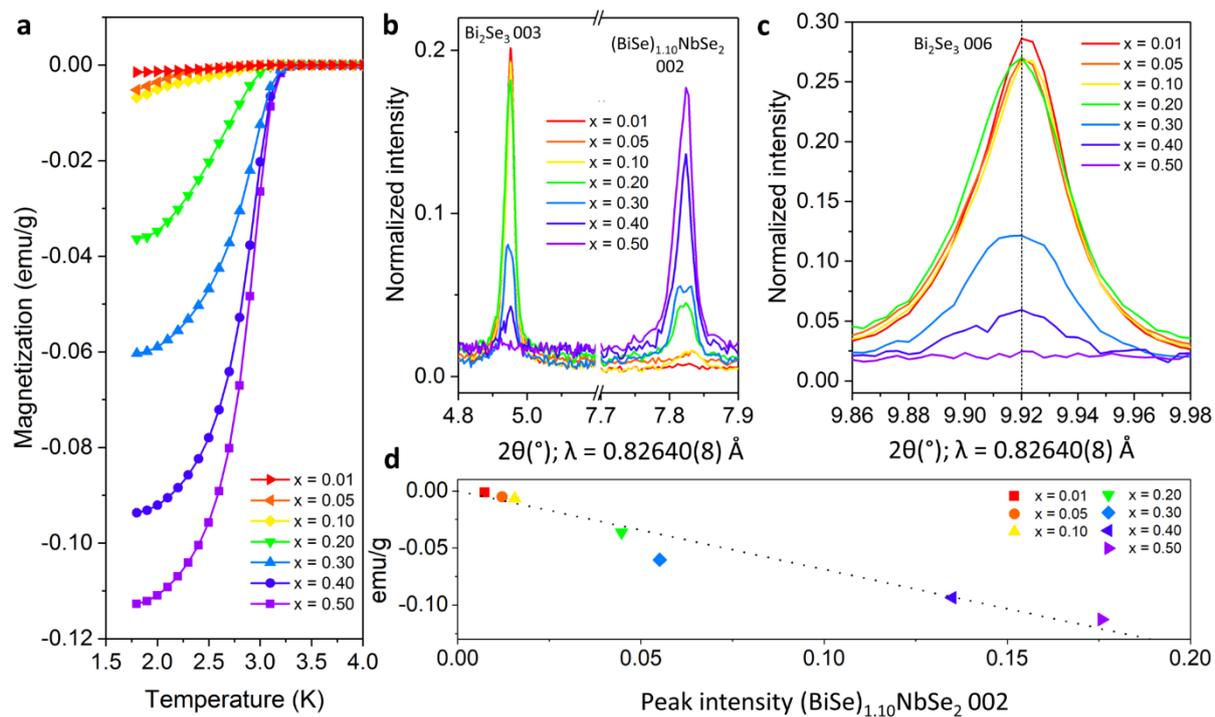

**Fig. 2 Relation between magnetometry and diffraction in Nb$_x$Bi$_2$Se$_3$. a** Magnetometry results (ZFC: zero field cooled) for Nb$_x$Bi$_2$Se$_3$ with varying Nb content. A full plot also containing the FC (field cooled) data is shown in Supplementary Fig. S1. **b** Selected peaks of the corresponding diffraction patterns indicating the relative intensities of the Bi$_2$Se$_3$ and (BiSe)$_{1.10}$NbSe$_2$ phases for varying Nb content. **c** Normalized intensity of the Bi$_2$Se$_3$ 006 peak with varying $x$. **d** Relation between the peak intensity of the 002 peak of (BiSe)$_{1.10}$NbSe$_2$ (from Fig. **b**) and magnetization at 1.8 K (from Fig. **a**)) with varying Nb content of the samples.

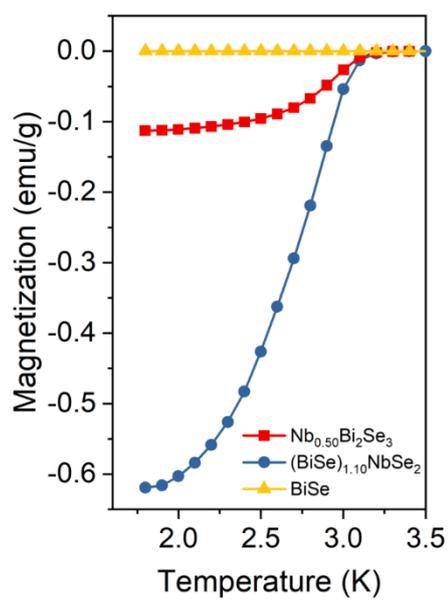

**Fig. 3 Magnetometry results (ZFC) for Nb$_{0.50}$Bi$_2$Se$_3$.** Nb$_{0.50}$Bi$_2$Se$_3$ (a mixture of solely BiSe and (BiSe)$_{1.10}$NbSe$_2$, as shown in Fig. 2 and Supplementary Fig. S4), BiSe and the pure misfit phase (BiSe)$_{1.10}$NbSe$_2$. A full plot also containing the FC data is shown in Supplementary Fig. S2.

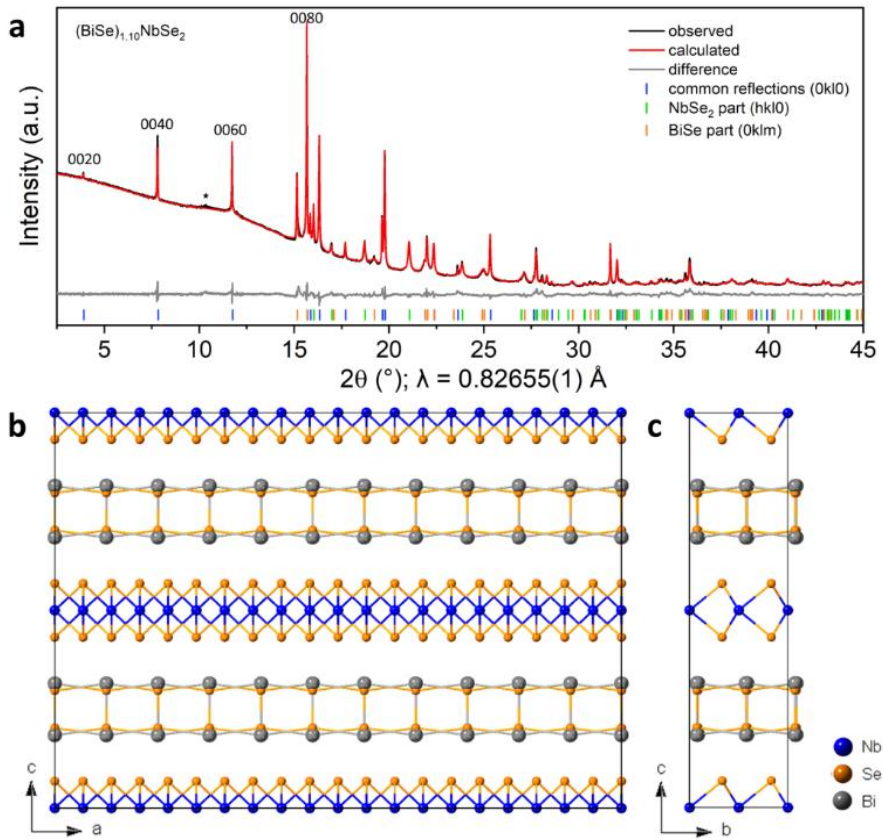

**Fig. 4 Diffraction pattern and crystal structure of (BiSe)$_{1.10}$NbSe$_2$. a** Rietveld fit to synchrotron data of (BiSe)$_{1.10}$NbSe$_2$. All peaks can be indexed with solely the main reflections of both sublattice. A few 00$l$0 peaks are indicated for clarity. The star symbol represents a tiny amount of BiSe impurity. **b,c** Average crystal structure of (BiSe)$_{1.10}$NbSe$_2$. The average structure drawn with anisotropic displacement ellipsoids is shown in Supplementary Fig. S7.

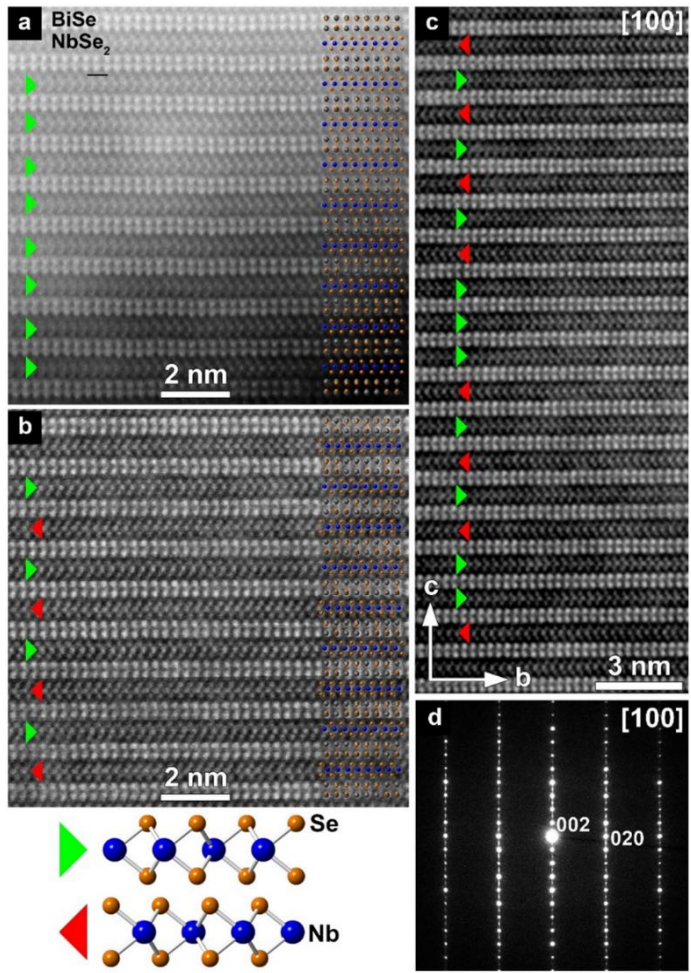

**Fig. 5 [100] HAADF-STEM images and electron diffraction pattern of (BiSe)$_{1.10}$NbSe$_2$.** The HAADF-STEM images clearly show a layered structure with alternating BiSe (appear brighter due to $Z_{Bi}$ = 83) and NbSe$_2$ (appear weaker due to $Z_{Nb}$ = 41) blocks. Structure overlay with **a** demonstrating all identical NbSe$_2$ layers and **b** mirrored NbSe$_2$ layers the *c*-direction. **c** HAADF-STEM image from a larger area showing a random distribution of the NbSe$_2$ layers. **d** Corresponding electron diffraction pattern showing elongated reflections 0*kl*: *l* = 2*n*+1 due to the stacking faults.

# Supplementary Information

# Misfit phase (BiSe)$_{1.10}$NbSe$_2$ as the origin of superconductivity in nobium-doped bismuth selenide

Machteld E. Kamminga,[a] Maria Batuk,[b] Joke Hadermann[b] and Simon J. Clarke[a*]

[a]Department of Chemistry, Inorganic Chemistry Laboratory, University of Oxford, South Parks Road, Oxford OX1 3QR, United Kingdom.

[b]Electron Microscopy for Materials Science (EMAT), University of Antwerp, Groenenborgerlaan 171, B-2020 Antwerp, Belgium.

[*] E-mail: simon.clarke@chem.ox.ac.uk

**Magnetometry data**

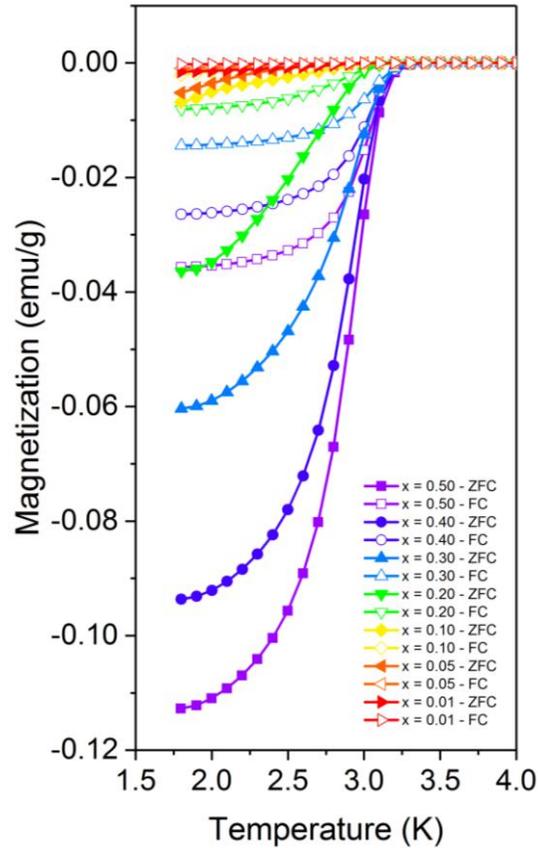

Fig. S1 Magnetometry results (ZFC and FC) for Nb$_x$Bi$_2$Se$_3$ with varying Nb content.

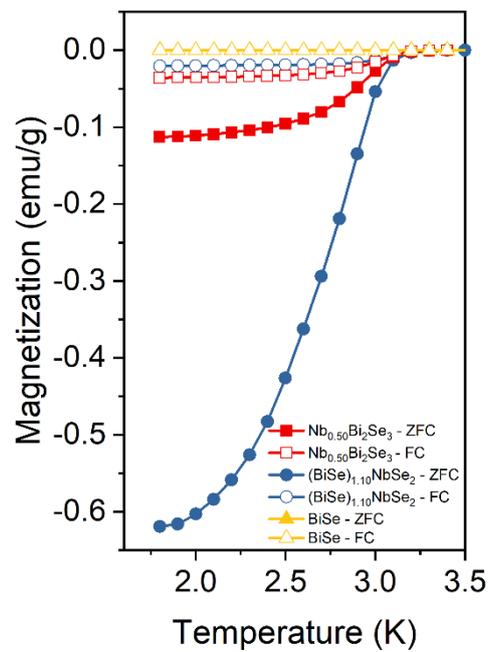

Fig. S2 Magnetometry results (ZFC and FC) for 'Nb$_{0.50}$Bi$_2$Se$_3$'. 'Nb$_{0.50}$Bi$_2$Se$_3$' (a mixture of solely BiSe and (BiSe)$_{1.10}$NbSe$_2$, as shown in Fig. 2), BiSe and the pure misfit phase (BiSe)$_{1.10}$NbSe$_2$.

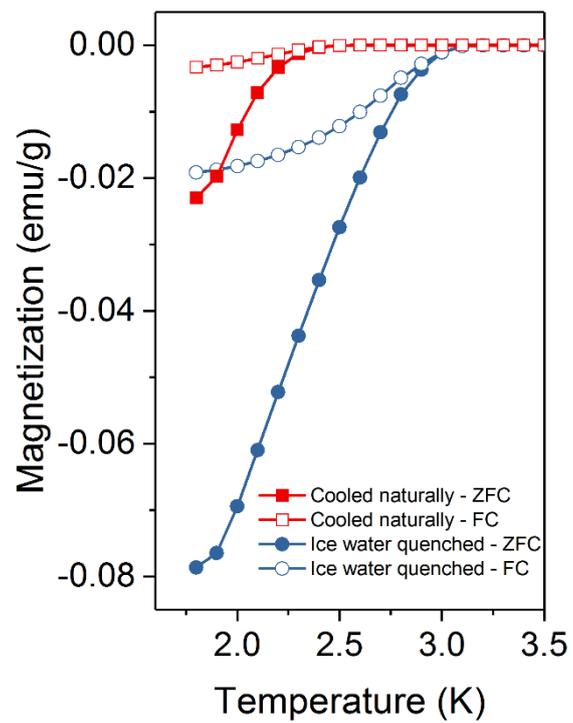

**Fig. S3** Magnetometry data of naturally cooled and ice/water quenched pure misfit samples.

**Supplementary Section 1: Investigation of the relationship between the superconducting volume fraction and the ratio of the $(BiSe)_{1.10}NbSe_2$ and BiSe in $Nb_{0.50}Bi_2Se_3$**

As can be seen in Fig. 2**b**, $Nb_{0.50}Bi_2Se_3$ contains no phase with the $Bi_2Se_3$ structure and therefore solely consists of the misfit phase and BiSe. *i.e.*

"$Nb_{0.50}Bi_2Se_3$" = 0.5 Nb + 2 Bi + 3 Se = 0.5 $(BiSe)_{1.10}NbSe_2$ + 1.45 BiSe

This is also supported by the EBSD and EDX data shown in Fig. S4. The EBSD image clearly shows crystallites with different backscatter intensities, caused by the relatively high Z value of BiSe compared to the $(BiSe)_{1.10}NbSe_2$.

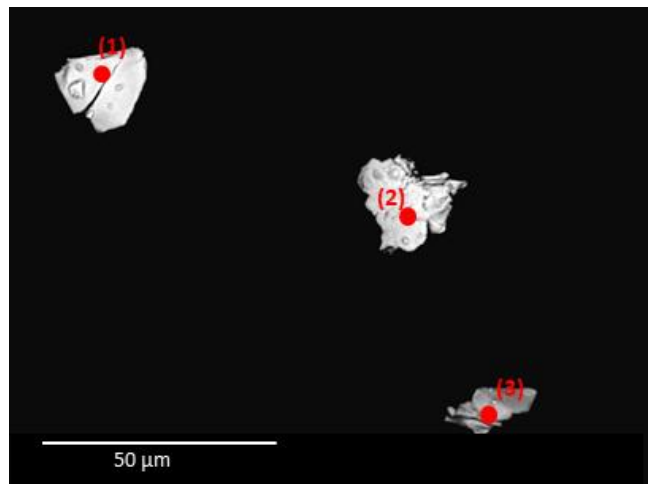

**Fig. S4 EBSD image of $Nb_{0.50}Bi_2Se_3$ showing two types of crystallites with different average Z value.** Point (1) and (2) have elemental ratios of ~1 Bi : ~1.1 Se and ~1 Bi : ~1.2 Se, respectively, which resemble BiSe, whereas point (3) has an elemental ratio of ~1 Nb : ~1.2 Bi : ~2.8 Se, which resembles $(BiSe)_{1.10}NbSe_2$.

After synthesis, we loosely ground the sample to obtain a coarse powder and sieved it to obtain four different batches using sieves with mesh sizes of 250 µm, 125 µm and 63µm. As shown in Fig. S5**a**, the sieved batches consist of different ratios of the $(BiSe)_{1.10}NbSe_2$ and BiSe phases. As shown in Fig. S5**b**, the different ratios of its constituents give rise to different superconducting volume fractions. Figure S5**c** shows the direct relationship between the superconducting volume fraction and the ratio of the two peaks and hence the ratio of the misfit phase and BiSe. As the superconducting volume fraction increase with the $(BiSe)_{1.10}NbSe_2$/BiSe ratio, we conclude that the misfit phase is indeed responsible for the superconductive properties of the compound and BiSe is not. This is confirmed by the data in Fig. 3 in the main article.

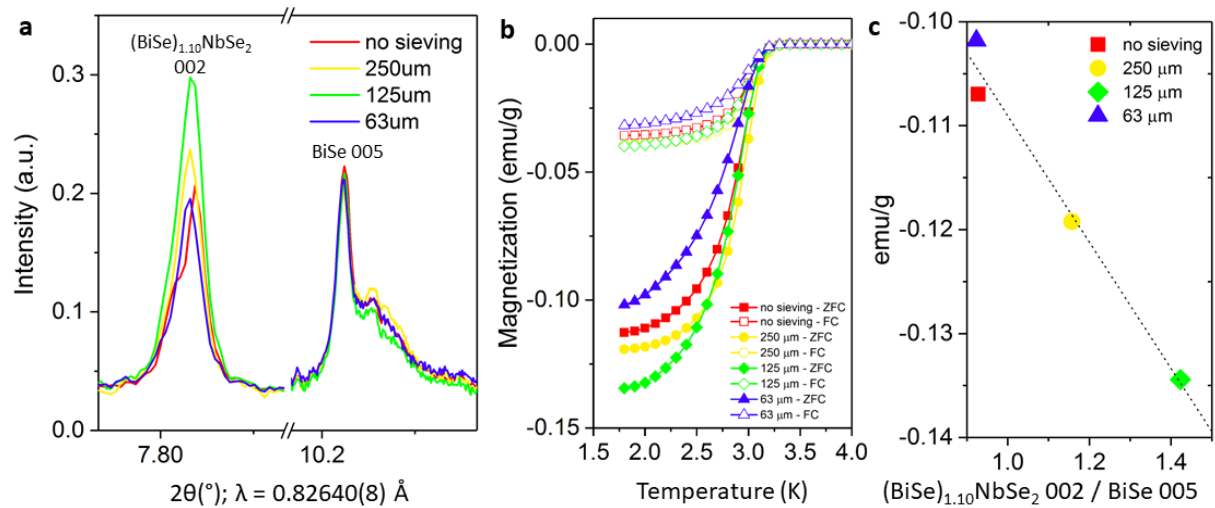

**Fig. S5 Relation between magnetometry and diffraction in $Nb_{0.50}Bi_2Se_3$ using differently sized sieves. a** Selected peaks of the diffraction patterns for $Nb_{0.50}Bi_2Se_3$ indicating the relative intensities of the $(BiSe)_{1.10}NbSe_2$ and BiSe phases. **b** Magnetometry results for $Nb_{0.50}Bi_2Se_3$ using differently sized sieves. These data correspond to the diffraction patterns of Fig. **a**. **c** Relation between the ratio of peak intensities of the 002 peak of the $(BiSe)_{1.10}NbSe_2$ phase and the 005 peak of the BiSe phase (from Fig. **a**) and magnetization at 1.8 K (from Fig. **b**) for $Nb_{0.50}Bi_2Se_3$ using differently sized sieves.

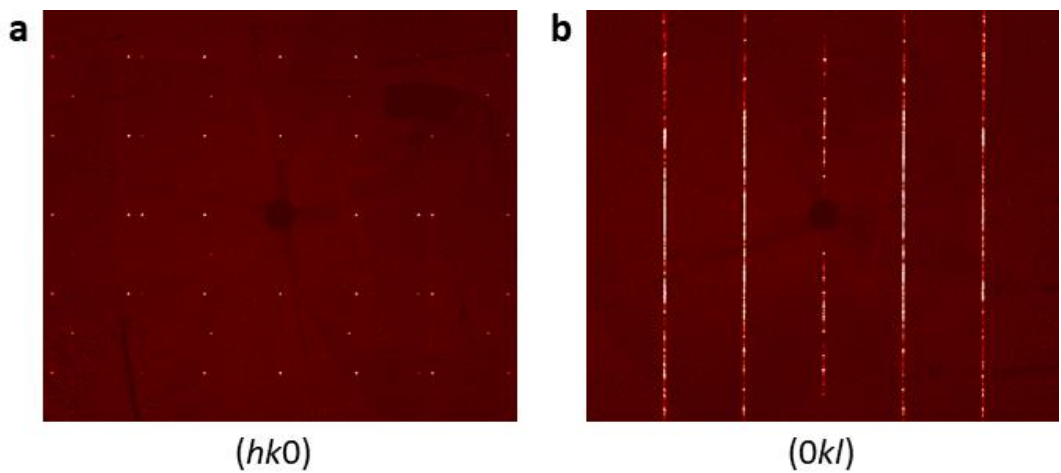

**Fig. S6 Single crystal reciprocal lattice planes of (BiSe)$_{1.10}$NbSe$_2$. a** ($hk0$) and **b** ($0kl$) reciprocal lattice planes of (BiSe)$_{1.10}$NbSe$_2$ obtained by making synthesized precession images from single crystal diffraction experiments showing severe stacking faults parallel to $c$.

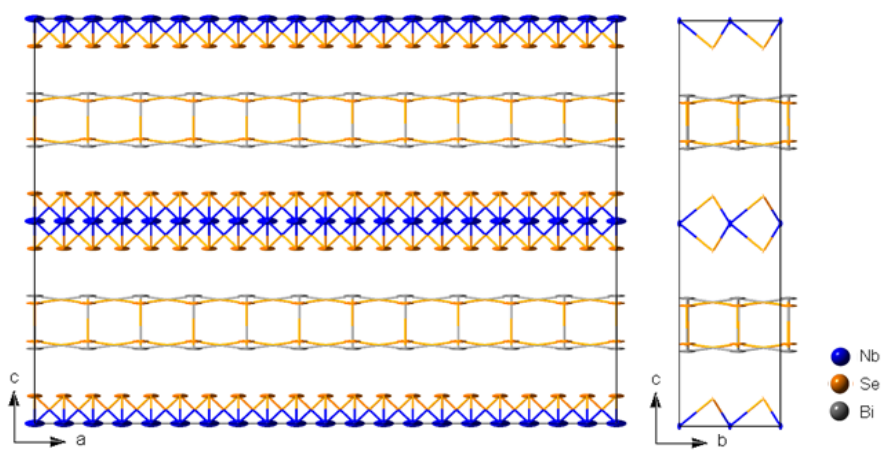

**Fig. S7 Average crystal structure of (BiSe)$_{1.10}$NbSe$_2$.** Anisotropic displacement ellipsoids of 50% probability are drawn. Note that the largest uncertainties are in the $a$ direction, which is the misfit direction.

**Supplementary Section 2: Rietveld refinement and modelling of the average crystal structure of (BiSe)$_{1.10}$NbSe$_2$ from synchrotron powder data and electron diffraction data**

In order to investigate the misfit phase (BiSe)$_{1.10}$NbSe$_2$ in more detail we synthesized the pure phase from the elements using a vapour transport synthesis method.[1] These so-called inorganic misfit layer structures are a special type of layered compound.[2] Contrary to periodic layered compounds (*e.g.* graphite or NbSe$_2$), intergrowth compounds are made up of two or more interpenetrating sublattices of different chemical composition. As each of these types of layers has different intralayer lattice constants, which generally do not match, it is not possible to define a single three-dimensional unit cell and space group to describe the entire system. Therefore, each component will have its own translational symmetry,[3] and each atom in the structure belongs to just one of the subsystems. Note that the lattice of one subsystem is generally incommensurate with the lattice of the other subsystem, ensuring the lack of three-dimensional translation symmetry of the complete system. In this description the crystal is described as consisting of two separate translationally symmetric subsystems. This only an approximation to the real crystal structure and is called the average structure: due to their mutual interaction, the subsystems in the real crystal will be modulated.[4] These incommensurate inorganic misfit layer compounds are widely represented by compounds with the general formula (MX)$_n$TX$_2$ (M = main group metal; X = chalcogenide; T = transition metal) and are characterized by the alternate stacking of two different types of layers.[2,5,6] The first type is a three-atom-thick layer of composition TX$_2$ with a structure analogous to the individual layers in either NbS$_2$ (edge-shared triangular prisms) or TiS$_2$ (edge-shared octahedra), typically denoted as the H sublattice. For NbSe$_2$ this means that Nb is in a triangular prismatic coordination by Se. The second type of layer has a structure corresponding to a two-atom thick (100) slice of a rock-salt-type structure of composition MX, typically denoted as the Q sublattice. The misfit parameter $n$ is defined as: $n = Z_Q V_H / Z_H V_Q$, with $Z$ the formula unit and $V$ the volume of each sublattice.[7] Examples of these misfit layer compounds are (PbS)$_{1.14}$NbS$_2$,[8] (LaS)$_{1.14}$NbS$_2$,[8] (SnS)$_{1.17}$NbS$_2$,[3,9] (BiS)$_{1.11}$NbS$_2$,[10,11] (BiS)$_{1.08}$TaS$_2$[12] and (BiSe)$_{1.09}$TaSe$_2$.[1,13]

Figure S8 represents electron diffraction patterns obtained from different parts of the crystal oriented along the [001] direction. The patterns consist of two sets of basic reflections, corresponding to the BiSe sublattice (indicated with subscript Q; pseudo-square pattern shown by pink squares) and to the NbSe$_2$ sublattice (subscript H; pseudo-hexagonal pattern shown by green circles).[7,14] The approximate unit cell parameters for the BiSe sublattice are $a_Q \approx 6$ Å and $b_Q \approx 6$ Å, for the NbSe$_2$ sublattice $a_H \approx 3.3$ Å and $b_H \approx 6$ Å. Note that $b_Q = b_H = b$, as shown by the shared 020 reflection (*i.e.* the layers are commensurate in this direction). The misfit ratio between the two sublattices is described by $a_H/a_Q$ and results in the misfit modulation vector $q_H \approx 0.55$. Additional weaker satellite reflections are evident in Fig. S8**b** that can be indexed using the modulation vector $\boldsymbol{q}_1 \approx (1/6)\, a_Q$. The length and orientation of this modulation vector $\boldsymbol{q}_1$ is in agreement with the modulation previously reported for the BiSe layer,[1,13] and will be discussed in more detail below. Note that the weak satellite reflections corresponding to the modulation vector $\boldsymbol{q}_1$ only become visible while using the beam stop, as the patterns in Fig. S8**a** and S8**b** were acquired from the same area without and with the beam stop, respectively. The electron diffraction pattern in Fig. S8**c** shows additional satellite reflections that can be indexed using the modulation vector $\boldsymbol{q}_2 \approx (1/10)\, b$, whereas $\boldsymbol{q}_1$ cannot be seen due to its low intensity in comparison to the central beam. In the enlarged cut around the 020 and 040 reflections, $\boldsymbol{q}_1$ and $\boldsymbol{q}_2$ are indicated.

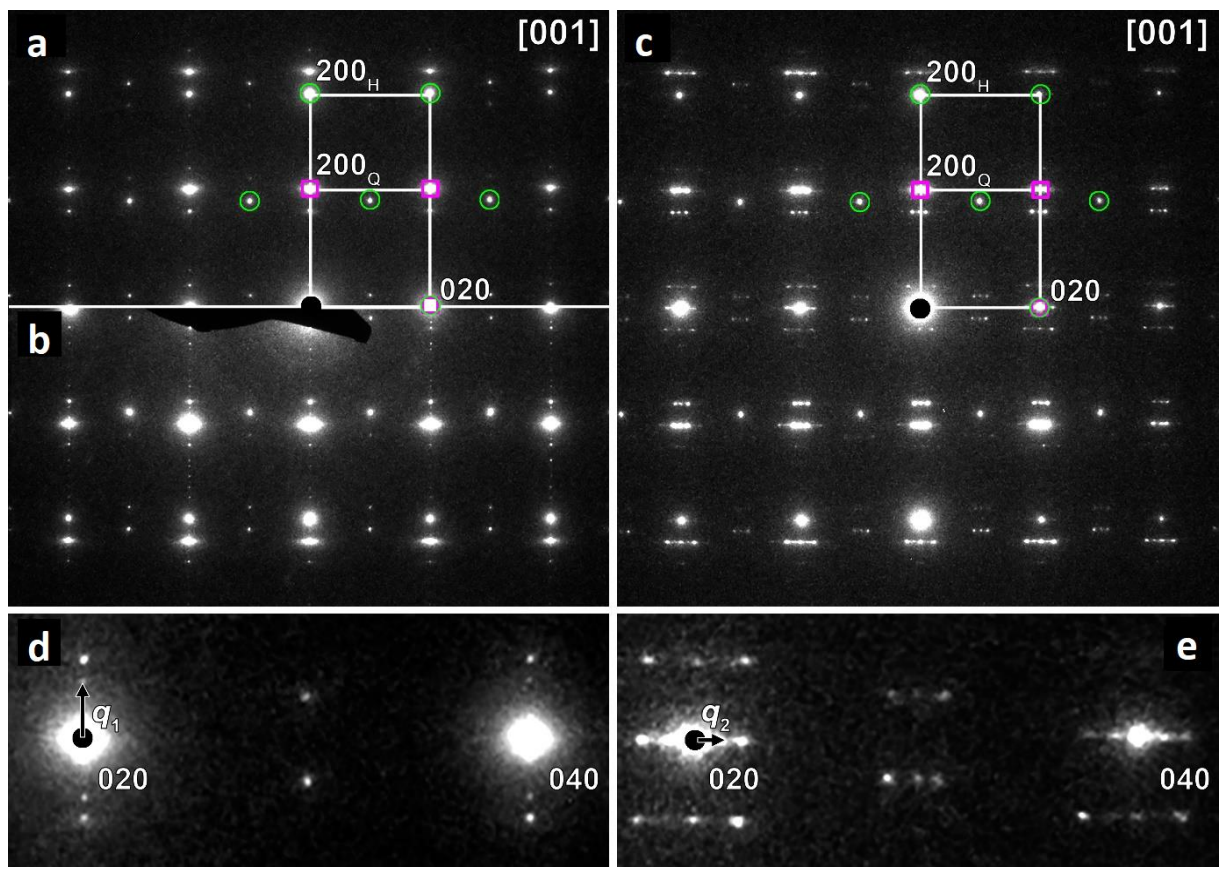

**Fig. S8 Electron diffraction patterns of (BiSe)$_{1.10}$NbSe$_2$.** Patterns acquired from the same area **a** without and **b** with the beam stop. **c** Electron diffraction pattern taken from another area. On the patterns the main reflections of the two basic sub-lattices are indicated with pink squares for the Q sub-lattice (BiSe) and green circles for H sub-lattice (NbSe$_2$). **d,e** Enlarged fragments from **b** and **c**, respectively, around the 020 and 040 reflections showing the satelite reflections: $q_1$ for **d** and $q_2$ for **e**.

Figure S9 shows the detailed indexation of the [001] electron diffraction pattern. For clarity, on the pattern acquired with a central beam stop a new virtual central beam is selected (indicated with black circle and indexed as 0000). White lines correspond to the main reflections of the misfit structure. Yellow lines indicate the satellites which cannot be indexed using the misfit lattice vectors and require $q_1 \approx (1/6)\, a_Q$ modulation vector. For clarity, indexes for main and satellite reflections are shown on two separate figures **a** and **b**, respectively.

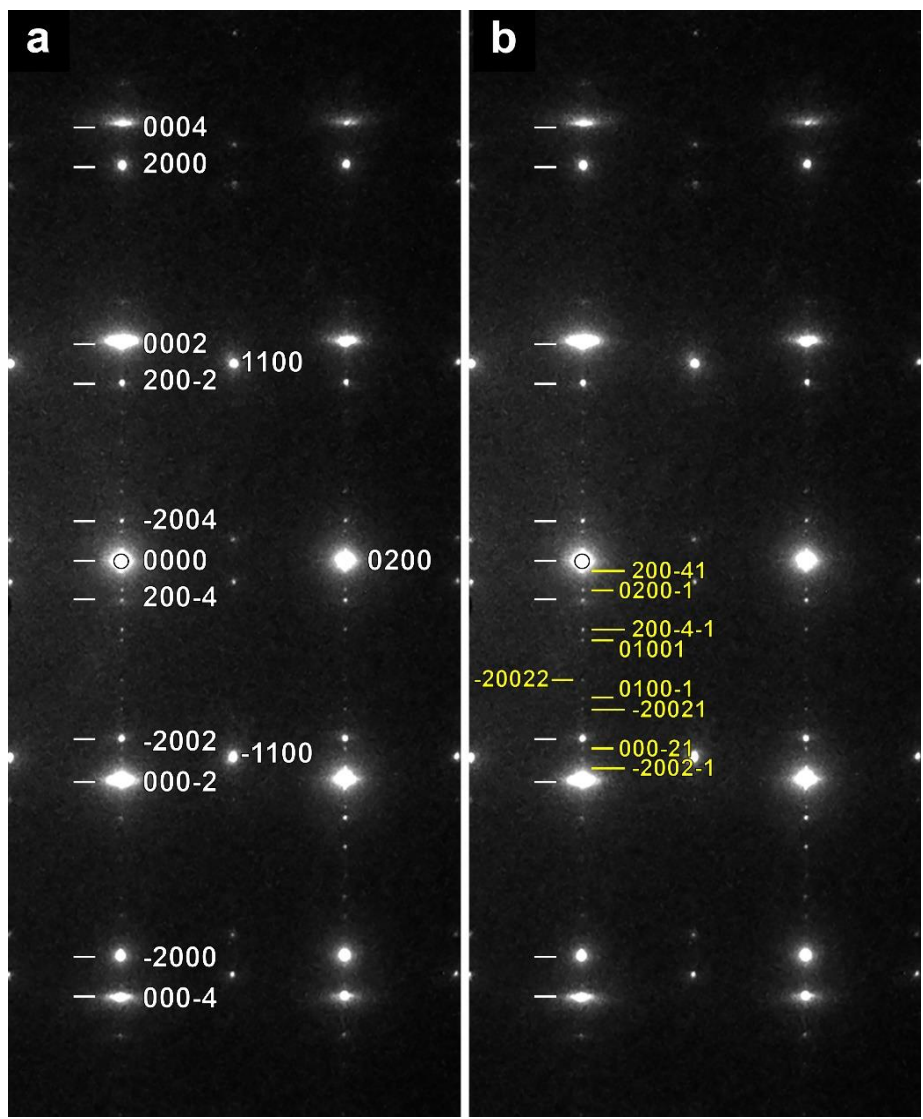

**Fig. S9 Fragments of the [001] electron diffraction patterns**. The patterns are acquired with a central beam stop and show satellite reflections. The indexes of the main reflections are shown in **a** and the indexes of the satellite reflections are shown in **b**. As the relevant reflections were hidden in the central part of the beam stop, we shifted the indexation grid to a clear area. The reflection indicated here as 0000 is in fact 0-400. Therefore, the real indices would be all given indices with -4 subtracted for the *k* index.

As discussed above, (BiSe)$_{1.10}$NbSe$_2$ can be described as simply consisting of two independent subsystems: NbSe$_2$ (H) and BiSe (Q). However, a better description of misfit compounds is provided by the superspace formalism.[3,15–17] Here, the complete structure is described in a (3 + *d*)-dimensional space, with the number of dimensions equal to the number of independent reciprocal vectors; here *d* =

1, when first considering only the misfit character. The symmetry is given by a single $(3 + d)$-dimensional space group, from which the subsystem symmetries can be derived.[4] This superspace formalism not only reduces the number of parameters used in refinement, but also allows the modulations of both subsystems to be described. For $(BiSe)_{1.10}NbSe_2$, the relation between the two sublattices H ($NbSe_2$) and Q (BiSe) is described by the interlattice matrix $W$ as shown in Table S1. $W^H$ is chosen as the unity matrix, whereas $W^Q$ describes how Q relates to H. The $W$ matrices can be thought of as defining a coordinate transformation in superspace. In particular, for the reflection indices it follows that

$$(H\ K\ L\ M) = (h_Q\ k_Q\ l_Q\ m_Q)W^Q = (h_H\ k_H\ l_H\ m_H)W^H$$

where $HKLM$ are the indices with respect to complete diffraction pattern in superspace and $h_Q k_Q l_Q m_Q$ and $h_H k_H l_H m_H$ are the reflection indices with respect to each sublattice. Note that $h_H k_H l_H m_H$ and $HKLM$ are identical as $W^H$ is chosen as the unity matrix. The $HKLM$ indices of the complete pattern, as shown in Fig. 4**a** of the main text, give rise to the following extinction rules: $H + K = 2n$, $H + L = 2n$ and $K + L = 2n$, indicating that the overall symmetry of the system is $F$-centred. This corresponds to the set of centring translations as listed in Table S1. In Fig. 4**a** of the main text, we fitted the pattern with superspace group $Fm2m(\alpha 00)000$ and the parameters listed in Table S1. Consequently, the subsystem superspace group for H is $Fm2m(\alpha 00)000$ and for Q is $Xm2m(\alpha 00)000$, corresponding to the modified set of centring translations as listed in Table S1. Note that the misfit parameter $n$ is given by $Z_Q V_H / Z_H V_Q = Z_Q a_H / Z_H a_Q \approx 1.10$, leading to the structural formula $(BiSe)_{1.10}NbSe_2$.

**Table S1 Structural parameters for (BiSe)$_{1.10}$NbSe$_2$.**

| Composite part H: NbSe$_2$ | Composite part Q: BiSe |
|---|---|
| $W^H = \begin{bmatrix} 1 & 0 & 0 & 0 \\ 0 & 1 & 0 & 0 \\ 0 & 0 & 1 & 0 \\ 0 & 0 & 0 & 1 \end{bmatrix}$ | $W^Q = \begin{bmatrix} 0 & 0 & 0 & 1 \\ 0 & 1 & 0 & 0 \\ 0 & 0 & 1 & 0 \\ 1 & 0 & 0 & 0 \end{bmatrix}$ |
| $a_H$ = 3.4398(1) Å  $b_H$ = 5.9887(1) Å  $c_H$ = 24.2165(1) Å | $a_Q$ = 6.2605(1) Å  $b_Q$ = 5.9887(1) Å  $c_Q$ = 24.2165(1) Å |
| $q_H = (a_H/a_Q\ 0\ 0)$ = (0.5494(1) 0 0) | $q_Q = (a_Q/a_H\ 0\ 0)$ = (1.8202(1) 0 0) |
| Z = 4 | Z = 8 |
| Superspace group: $Fm2m(\alpha00)(000)$ (no. 247) | Superspace group: $Xm2m(\alpha00)(000)$ (no. 247) |
| Centering vectors: (0 0 0 0) (0 ½ ½ 0) (½ 0 ½ 0) (½ ½ 0 0) | Centering vectors: (0 0 0 0) (0 ½ ½ 0) (0 0 ½ ½) (0 ½ 0 ½) |
| Symmetry operators: $x_1\ x_2\ x_3\ x_4$ $-x_1\ x_2\ -x_3\ -x_4$ $-x_1\ x_2\ x_3\ -x_4$ $x_1\ x_2\ -x_3\ x_4$ | Symmetry operators: $x_1\ x_2\ x_3\ x_4$ $-x_1\ x_2\ -x_3\ -x_4$ $-x_1\ x_2\ x_3\ -x_4$ $x_1\ x_2\ -x_3\ x_4$ |

In Fig. **4a** of the main text, we fitted the pattern with superspace group $Fm2m(\alpha00)000$ and the parameters listed in Table S1. As shown by the difference curve, the fit gives a reasonable fit to the data. Unfortunately, indexing satellite peaks is challenging with powder data. In fact, in the majority of similar cases only first order satellites are visible on conventional powder X-ray diffraction patterns.[18] For (BiSe)$_{1.10}$NbSe$_2$, no satellites were observed, not for $q_H$, $q_1$ or $q_2$, and the full pattern can be indexed with solely the main reflections of both sublattices. Therefore, it was not possible to solve the modulation, but we could describe the average structure to good agreement.

Figures **4b,c** of the main text show the average crystal structure of (BiSe)$_{1.10}$NbSe$_2$ as obtained by the Rietveld refinement form Fig. **4a**. According to literature, the average structure of a single slab of BiX (X = S, Se) in bismuth-containing compounds is the same as in other misfit compounds,[12,19]

wherein the BiX layer has atomic sites that on the average length scale probed by diffraction methods are occupied by equal amounts of Bi and X in a disordered way. However, the real ordered structure appears different.[13] Instead of ordering on an $a \times b$ lattice (with $a \approx b \approx 6$Å), observed for all other compounds, the unit cell of the BiX layers is found to be $6a \times b$.[13] The ordering in $(BiSe)_{1.09}TaSe_2$ is such that four $Bi_2$ pairs occur in each cell, accompanied by four $Se_2$ pairs with nonbonding distances.[1] As stated above, no satellites were observed in our data, so we focussed on modelling the average structure of our misfit compound. In this model, we set the occupancies of Bi and Se in the BiSe slab to ½ and put constraints in place to make their anisotropic thermal parameters as well as their $y$ parameters equal, as they occupy the same site. The structure coordinates are listed in Table S2.

**Table S2 Basic structure coordinates for $(BiSe)_{1.10}NbSe_2$.**

|  | lattice | $x_1$ | $x_2$ | $x_3$ |
|---|---|---|---|---|
| Nb1 | H | 0 | 0.0010(12) | 0 |
| Se1 | H | 0.5 | -0.1674(12) | -0.0680(5) |
| Bi2 | Q | 0 | 0.074(3)[a] | 0.3153(6)[c] |
| Se2 | Q | 0 | 0.074(3)[a] | 0.2972(12)[d] |
| Bi3 | Q | 0.5 | 0.086(2)[b] | 0.3153(6)[c] |
| Se3 | Q | 0.5 | 0.086(2)[b] | 0.2972(12)[d] |

[a,b,c,d] Fixed to be identical in each case

**Supplementary Section 3: STEM results**

The elemental composition of the misfit compound was determined from STEM-EDX maps taken from 10 different areas of the FIB slices. Atomic resolution STEM-EDX maps acquired along the [100] zone, showing the layered arrangement of Bi, Nb and Se in the structure are shown in Fig. S10.

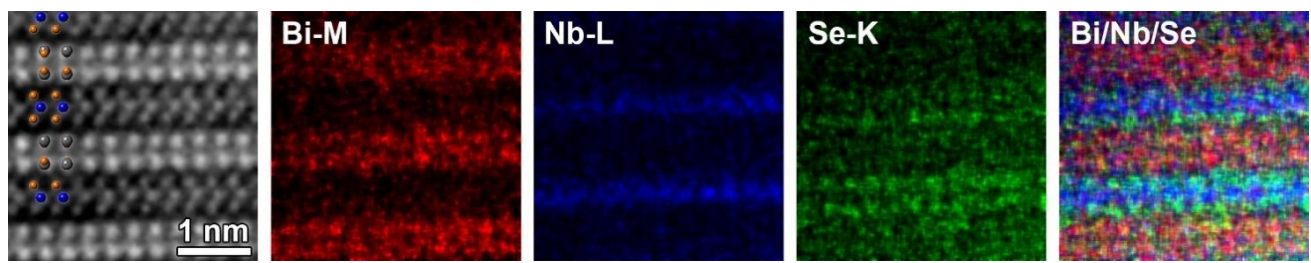

**Fig. S10 HAADF-STEM image of (BiSe)$_{1.10}$NbSe$_2$ with atomic resolution and corresponding STEM-EDX maps.** The STEM-EDS map show the presence of Bi, Se and Nb in the structure.

As shown in Fig. **4c** of the main text, the average structure as refined from the powder X-ray diffraction experiment consists of NbSe$_2$ slabs that all point in the same direction, and the HAADF-STEM image in Fig. **5a** which probes shorter range structural features show such a region in which the NbSe$_2$ slabs are all similarly aligned. As can be seen from the HAADF-STEM images obtained at a larger scale, there are NbSe$_2$ layers where the NbSe$_2$ prisms look mirror related in respect to the *ac* plane. Within the same NbSe$_2$ layer the orientation of the prisms remain constant and do not switch. On average, the distribution of such layers in the crystal is random. However, regions with perfect alternation were observed as well (Fig. **5b**). The alternating, or random orientations would lead to violation of the *HKLM*: $H + L = 2n$ and *HKLM*: $K + L = 2n$ reflection conditions for *F*-centring of the NbSe$_2$ lattice, hence the 00$L$0 reflections with $L$ odd would no longer be systematically absent, and they are indeed clearly evident in the electron diffractogram shown in Fig. **5d** of the main text. Furthermore, the streaking found in the single crystal diffraction pattern (Fig. **S6b**) shows evidence on the length scale probed by X-ray diffraction for random orientations of the NbSe$_2$ layers. In the powder X-ray diffraction pattern in Fig. **4a**, however, these reflections are not evident above the background level so we used the *F*-centred cell wherein all NbSe$_2$ layers point in the same direction to account for the powder data. An approximate commensurate model consisting of NbSe$_2$ slabs pointing in alternating directions as derived from TEM results is described below.

Analysis of the HAADF-STEM images taken along the [100] orientation allowed us to propose approximate commensurate models of the misfit compound with NbSe$_2$ prisms with alternating orientations along the *c*-direction (Figure 5**b** of the main text). We unveil two possibilities that match the experimental data, so-called, model1 and model2, which are different in the position of the prisms along the x-axis (Table S3). For these two models no difference can be observed neither on electron diffraction patterns nor on images taken along [100] or [010] directions (Figure S10). The simulated [100] electron diffraction patterns match the observed one with correct loss of the 0*kl*: *l* = 2*n* reflection condition and only 0*kl*: *k* = 2*n* remaining. When viewed along the [001] zone, the structures show small differences in the positions of the Nb and Se atomic columns. Unfortunately, on our [001] experimental HAADF-STEM images we can only clearly see the positions of the Bi atomic columns, whereas the Nb and Se atomic columns are smeared. On the simulated [001] electron diffraction patterns for the two models, only a small difference in intensities of some reflections can be noticed, which we do not see on the experimental patterns or Fourier transforms (FFT's) taken from the corresponding images. Therefore, we cannot eliminate either of the two models as a possibility.

**Table S3 Atomic parameters for (BiSe)$_{1.10}$NbSe$_2$ models with NbSe$_2$ prisms with alternating orientations along the *c*-direction.** The models are made in *P*1 space group with the following lattice parameters: *a* = 3.4392 Å, *b* = 5.9875 Å, *c* = 24.2127 Å, α = β = γ = 90°. Bi and Se atoms have 0.5 occupancy due to their random distribution in the average structure of (BiSe)$_{1.10}$NbSe$_2$ with the same NbSe$_2$ prisms (Table S2).

|       | occupancy | Model1 | | | | Model 2 | | |
|-------|-----------|---|---|---|---|---|---|---|
|       |           | x | y | z |   | x | y | z |
| Bi1   | 0.5 | 0 | 0.25 | 0.3157 | | 0 | 0.25 | 0.3157 |
| Bi2   | 0.5 | 0 | 0.75 | 0.8157 | | 0 | 0.75 | 0.8157 |
| Bi3   | 0.5 | 0 | 0.25 | 0.8157 | | 0 | 0.25 | 0.8157 |
| Bi4   | 0.5 | 0 | 0.75 | 0.3157 | | 0 | 0.75 | 0.3157 |
| Bi5   | 0.5 | 0 | 0.25 | 0.6843 | | 0 | 0.25 | 0.6843 |
| Bi6   | 0.5 | 0 | 0.75 | 0.1843 | | 0 | 0.75 | 0.1843 |
| Bi7   | 0.5 | 0 | 0.25 | 0.1843 | | 0 | 0.25 | 0.1843 |
| Bi8   | 0.5 | 0 | 0.75 | 0.6843 | | 0 | 0.75 | 0.6843 |
| Nb1   | 1 | 0 | 0.84 | 0 | | 0.5 | 0.84 | 0 |
| Nb2   | 1 | 0 | 0.66 | 0.5 | | 0 | 0.66 | 0.5 |
| Nb3   | 1 | 0.5 | 0.16 | 0.5 | | 0.5 | 0.16 | 0.5 |
| Nb4   | 1 | 0.5 | 0.34 | 0 | | 0 | 0.34 | 0 |
| Se1   | 1 | 0.5 | 0 | 0.0666 | | 0 | 0 | 0.0666 |
| Se2   | 1 | 1 | 0 | 0.5666 | | 1 | 0 | 0.5666 |
| Se3   | 1 | 0.5 | 0 | 0.9334 | | 0 | 0 | 0.9334 |
| Se4   | 1 | 1 | 0 | 0.4334 | | 1 | 0 | 0.4334 |
| Se5   | 1 | 0 | 0.5 | 0.9334 | | 0.5 | 0.5 | 0.9334 |
| Se6   | 1 | 1 | 0.5 | 0.0666 | | 0.5 | 0.5 | 0.0666 |
| Se7   | 1 | 0.5 | 0.5 | 0.5666 | | 0.5 | 0.5 | 0.5666 |
| Se8   | 1 | 0.5 | 0.5 | 0.4334 | | 0.5 | 0.5 | 0.4334 |
| Se9   | 0.5 | 0 | 0.25 | 0.2976 | | 0 | 0.25 | 0.2976 |
| Se10  | 0.5 | 0 | 0.75 | 0.7976 | | 0 | 0.75 | 0.7976 |
| Se11  | 0.5 | 0 | 0.25 | 0.7976 | | 0 | 0.25 | 0.7976 |
| Se12  | 0.5 | 0 | 0.75 | 0.2976 | | 0 | 0.75 | 0.2976 |
| Se13  | 0.5 | 0 | 0.25 | 0.7024 | | 0 | 0.25 | 0.7024 |
| Se14  | 0.5 | 0 | 0.75 | 0.2024 | | 0 | 0.75 | 0.2024 |
| Se15  | 0.5 | 0 | 0.25 | 0.2024 | | 0 | 0.25 | 0.2024 |
| Se16  | 0.5 | 0 | 0.75 | 0.7024 | | 0 | 0.75 | 0.7024 |

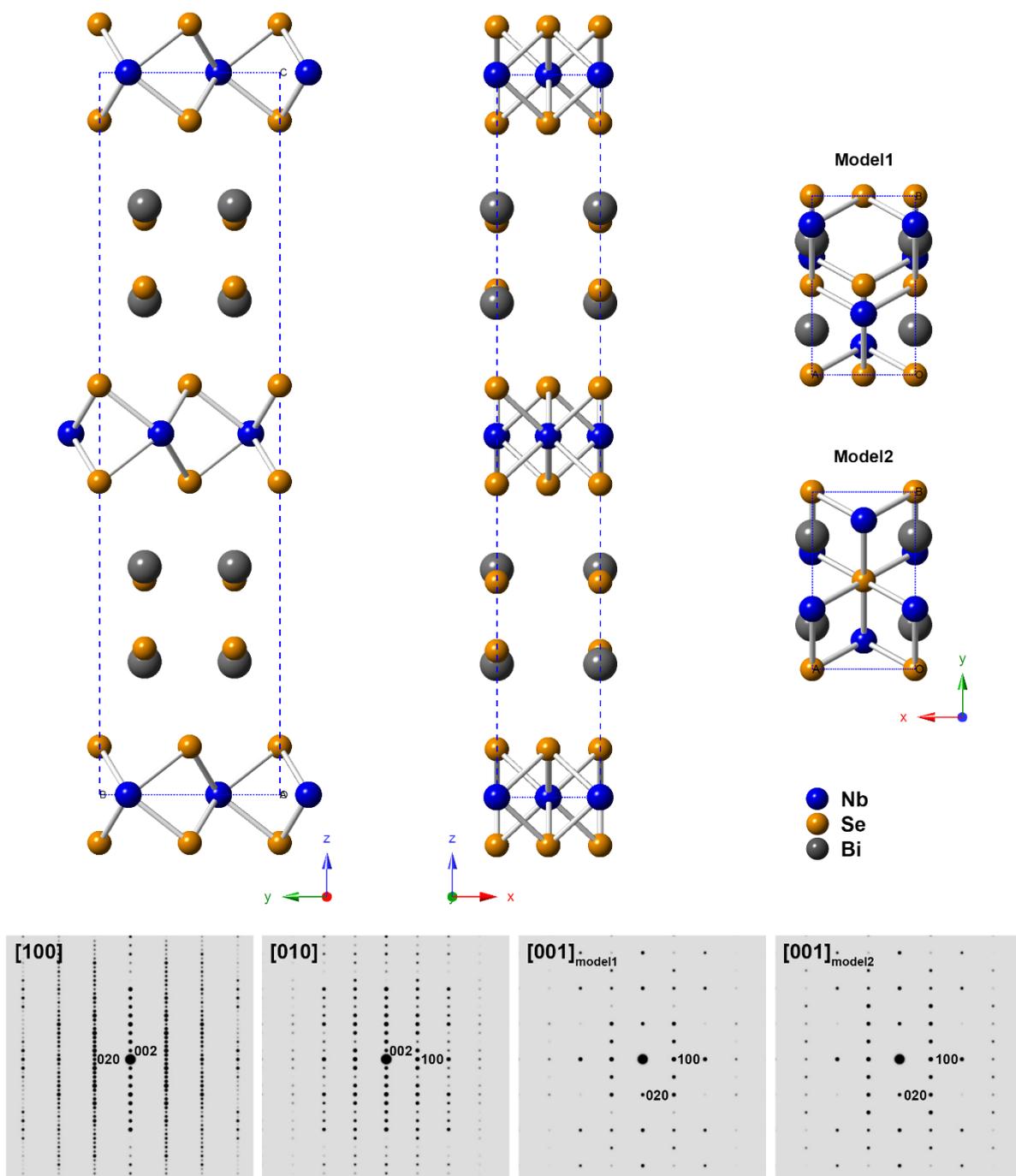

**Fig. S11 Approximate commensurate models of the misfit compound. Top**: Structural models of (BiSe)$_{1.10}$NbSe$_2$ with NbSe$_2$ prisms with alternating orientations along the *c*-direction. **Bottom:** Simulated electron diffraction patterns.

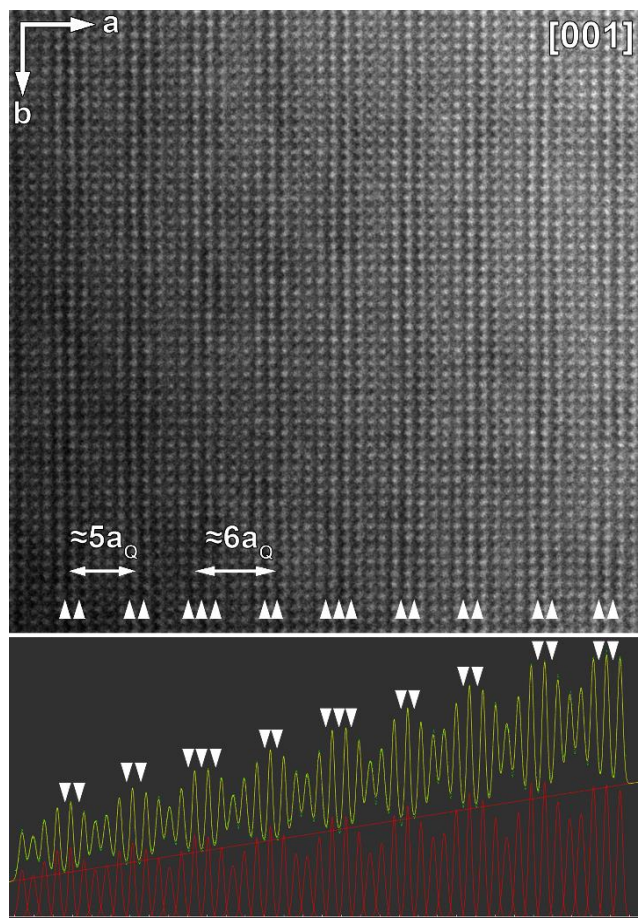

**Fig. S12 Modulation of the interatomic distances in (BiSe)$_{1.10}$NbSe$_2$. Top**: [001] HAADF-STEM image. Arrowheads point to some dark rows caused by the increase of the interatomic distance. Their presence result in a modulation with a periodicity of ~5-6 $a_Q$. **Bottom:** Corresponding intensity profile measured along 49 rows of Bi atomic columns. The measured profile was fitted by a set of Gaussian functions using the Fityk software.[20] Green – measured points, yellow – Gaussian functions, red – peaks with subtracted background. Arrows indicate interatomic distances longer than 3.4 Å.

A HAADF-STEM image acquired from the thin area oriented along the [001] direction is shown in Fig. S12. Darker rows parallel to the (100) planes are evident on the image (marked with arrowheads). These appear in an almost periodic fashion. These darker rows are caused by a modulation of the interatomic distances from the normal 3.0 Å to about 3.4-3.5 Å, as determined directly from the images. Their periodicity is approximately 5-6 $a_Q$ unit cells which agrees with the $q_1$ modulation vector determined from the electron diffraction patterns in Fig. S8. This modulation of the interatomic distances points towards a modulation of the BiSe block which is likely to be similar to that observed for $(BiSe)_{1.09}TaSe_2$ in which ordering on an $6a \times b$ lattice occurs in which four $Bi_2$ pairs are found in each cell, accompanied by four $Se_2$ pairs with nonbonding distances.[1,13] Note that the lack of observed satellites in the powder X-ray pattern in Fig. 4**a** in the main text prohibited refinement of this modulation in $(BiSe)_{1.10}NbSe_2$, resulting in the average structure of the BiSe slabs that have Bi and Se occupying the same site with occupancy of ½. Concerning the second modulation with vector $q_2 \approx (1/10)\ b$ observed on some electron diffraction patterns (Fig. S8**e**), we cannot pinpoint any structural features on the images responsible for this modulation and therefore its origin remains of interest for future research.